# EFIMOV STATES IN HALO NUCLEI

D.V. Fedorov[1], A.S. Jensen and K. Riisager

*Institute of Physics and Astronomy, Aarhus University*
*DK-8000 Aarhus C, Denmark*

**Abstract**

We investigate conditions for the occurrence of Efimov states in the recently discovered halo nuclei. These states could appear in systems, where one neutron is added to a pronounced one-neutron halo nucleus. Detailed calculations of the properties of the states are presented and promising candidates are discussed.

**Introduction.** The Efimov effect was discovered theoretically as a peculiar behavior of three particles interacting by short-range two-body potentials [1]. The effect appears when at least two of the binary subsystems have extremely large scattering lengths or bound states at nearly zero energy. Then a number of three-body bound states arises with enormous spatial extension and exceedingly small binding energies. Even when none of the binary subsystems are bound, the three-body system may have a huge number of bound states. One surprising consequence is that *more* three-body bound states can appear by *weakening* the potentials such that the scattering lengths increase.

These states have not been seen experimentally yet, but their possible existence have been investigated theoretically in a few systems like the three $^4$He-atoms ($^4$He-trimer) [2]. Such investigations have been rather demanding in terms of computer capacity as the extremely small binding energies and the large spatial extensions of the systems put strict requirements on the numerical accuracies. However, recently a new efficient method was developed for solving the Faddeev equations in coordinate space [3]. The method is particularly suitable for studying the Efimov effect.

Halo nuclei are nuclear systems with unusually large spatial extensions and small binding energies [4]. They are mainly found on the neutron dripline, but may also appear as excited states of normal nuclei as well as on the proton dripline. The characteristic properties of halo nuclei off hand matches the description of Efimov states rather well. The fact that they often are describable as two- or three-body systems furthermore improves the resemblance. The purpose of the present letter is to investigate the conditions for the occurrence of Efimov states in the new structures called halo nuclei. This is only possible to do accurately and systematically due to the newly developed method [3].

**The method.** The Faddeev equations in coordinate space are solved by use of an expansion in a complete set of generalized angular functions

$$\hat{\Psi} = \sum_n \rho^{-5/2} f_n(\rho) \hat{\Phi}_n(\rho, \Omega) , \qquad (1)$$

where $\hat{\Psi}$ and $\hat{\Phi}_n$ are three-component wavefunctions that depend on a set of Jacobi coordinates, i.e. an angular set denoted $\Omega$ and the effective radial coordinate $\rho$ (hyperradius) defined as $\rho^2 = \sum_{i=1}^{3} A_i r_i^2$ in terms of mass numbers $A_i$ and coordinates of the particles $\mathbf{r}_i$ in the c. m. frame. The functions $\hat{\Phi}_n$ are for each $\rho$ chosen as the eigenfunction of the angular part of the Faddeev equations [3]. The corresponding eigenvalue $\lambda_n(\rho)$ enters as an effective potential in the system of coupled radial equations

$$\left(-\frac{\partial^2}{\partial \rho^2} - \frac{2mE}{\hbar^2} + \frac{\lambda_n(\rho) + 15/4}{\rho^2}\right) f_n(\rho) = \sum_{n'} C_{nn'} f_{n'}(\rho) , \qquad (2)$$

where $m$ is the nucleon mass, $E$ is the total energy and the matrix elements of the non-diagonal (operator) terms are denoted by $C_{nn'}$. When at least two of the scattering lengths $a_i$ are infinite the lowest eigenvalue $\lambda$ will asymptotically approach a negative constant $\lambda^{(\infty)} < 0$, which gives rise to an attractive $\rho^{-2}$ potential with an infinite number of bound states. If the scattering lengths are large, but finite, $\lambda(\rho)$ remains close to $\lambda^{(\infty)}$ up to $\rho \approx \sum a_i$, where it starts to deviate and eventually either approaches zero as $-16 \sum a_i/(\pi\rho)$ (no binary bound state) or diverges parabolically as $-2m|E_x|\rho^2/\hbar^2$ (binary bound state of energy $E_x$) [3]. If only one of the three scattering lengths is infinitely large then the lowest $\lambda$ also approaches a negative constant although in this case the absolute value of the constant is too small to produce an infinite number of bound states.

---

[1]On leave from the Kurchatov Institute, 123182 Moscow, Russia

Long-range repulsive interactions hinder the effect. This is true in particular for the Coulomb potential in nuclei that gives a positive contribution to $\lambda(\rho \to \infty) \propto \rho$ so that the effective potential eventually will be repulsive at large distances. This does not mean that the Coulomb potential strictly exclude occurrence of these peculiar states, but there will only be a finite number of them. Relative angular momenta also lead to positive contributions to $\lambda(\rho \to \infty)$ and therefore also decreases the probability of finding these states. Thus the most promising nuclear candidates must consist of two neutrons in zero angular momentum states around a core nucleus. These three "particles" must behave like a three-body system and therefore at most only weakly involve the core degrees of freedom. These requirements are also the conditions for the occurrence of two-neutron halo systems.

**Calculations.** The neutron-neutron interaction is rather well known [5]. Since only the low energy properties are important [6], we shall use a simple attractive central potential of gaussian form, which reproduces the measured s-state scattering length and effective range, i.e. $-31\text{MeV}\exp(-(r/1.8\text{fm})^2)$ [7]. The neutron-core potential is also assumed to be gaussian, i.e. $-S_{cn} \exp(-(r/2.55\text{fm})^2)$. We shall use a notation where the selected binary subsystem is labeled by "x" and the last particle's motion relative to the center of mass of the "x"-system is labeled by "y". Spin, orbital- and total angular momentum are denoted, respectively by $S$, $L$ and $J$. Our neutron-core potential is spin independent and we therefore assume a spin-zero core. We shall furthermore consider positive parity and total angular momentum $J = 0$, which is supposed to be the most favorable conditions for finding the Efimov effect.

The neutron-core two-body system is first investigated as function of the strength $S_{cn}$. The scattering length is computed for each potential and the angular part of the three-body problem is solved. Eigenvalues $\lambda_n(\rho)$ and wavefunctions $\hat{\Phi}_n$ are obtained for $\rho$-values up to 20 times the largest scattering length. The resulting lowest lying eigenvalues are shown in fig. 1 as function of $\rho$ for a neutron-core potential with one slightly bound state with the binding energy $B_{cn} =$ 40 keV. We recognize the hyperharmonic spectrum at $\rho = 0$ as $\lambda_n = K(K + 4)$ where $K = 2n$ is a non-negative even integer [8]. The odd integers are connected with odd parity. At large $\rho$ the convergence towards the same spectrum is clearly seen for the curves number 3,4 and 6, whereas the curves number 1, 2 and 5 have just started to bend over respectively towards the parabolic $(-\rho^2)$ divergence, 0 and 32. The slow divergence of the lowest eigenvalue is the signature of a barely bound state in the neutron-core potential.

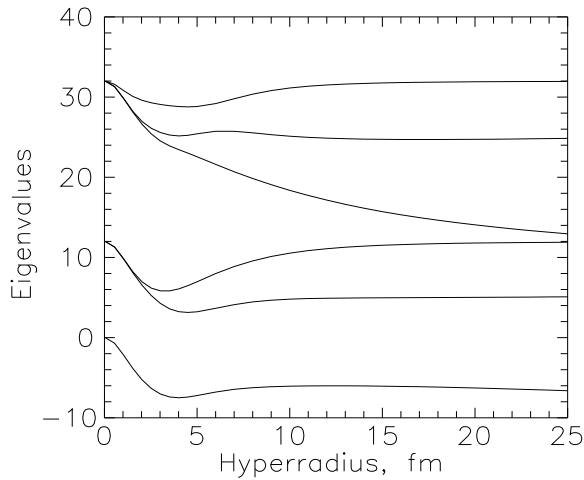

Figure 1: The six lowest lying angular eigenvalues $\lambda$ as function of $\rho$ for a neutron-core gaussian potential, strength $S_{cn}$ =10.7 MeV and range parameter 2.55 fm, corresponding to one slightly bound state of binding energy $B_{cn}$ =40 keV. The core mass is 9 times the nucleon mass.

The radial equation in eq. (2) is now solved and the resulting energies for the ground state and the first excited state are shown in fig. 2 as function $S_{cn}$. The two- and three-body systems are bound respectively for $S_{cn}$ >9.510 MeV or $S_{cn}$ >6.6 MeV. Extremely close to the two-body threshold, but still before ($S_{cn}$ ≈9.505 MeV), appears the first excited three-body state, where the neutron-core scattering length is about 18000 fm. Infinitely many bound states appear, one after the other, when $S_{cn}$ is increased from 9.505 MeV to 9.510 MeV. The closer we are to the two-body threshold the larger is the number of bound states. This narrow region could be called the Efimov region.

The convergence of the $\lambda$-expansion in eq. (1) is very fast for both ground and excited states. For the gaussian potentials the first $\lambda$ already provides good accuracy. The contribution, $\int |f_2(\rho)|^2 d\rho$, to the norm of the total wave function from the second $\lambda$ is about 0.1%. The second $\lambda$ changes the binding energy by about 2% compared to the result when only the lowest $\lambda$ is included.

The binding energy of the first excited state increases until the state disappears into the two-body continuum at $S_{cn} \approx 11.9$ MeV, where the neutron-core scattering length is about −15 fm. Therefore the one-neutron separation energy for this state ($B_1 - B_{cn}$) will first increase and then decrease as the potential strength is increased. Note that the states preferentially occur for bound neutron-core systems. The ground state wavefunction is concentrated in the pocket region near 4 fm. The excited state has unlike the ground state a node in the radial wavefunction, which is peaked at $\rho$-values far larger than the ranges of the nuclear potentials. Consequently the ground state and the first excited state

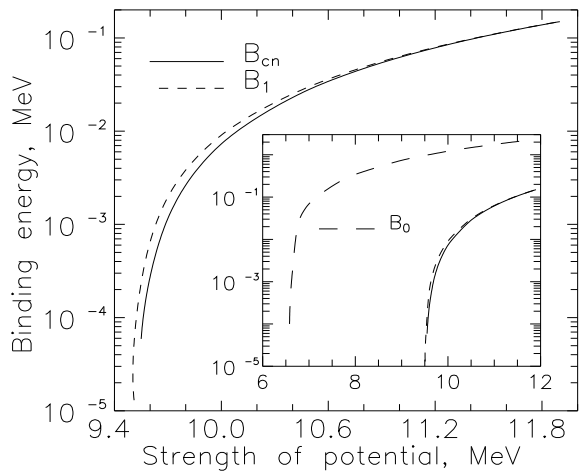

Figure 2: Binding energies of the two-body ground state ($B_{cn}$ solid curve) and the first excited state ($B_1$ dashed curve) of the three-body system as functions of potential strength $S_{cn}$. The insert also includes the three-body ground state ($B_0$ long-dashed curve). The core mass is 9 times the nucleon mass.

are orthogonal although they both have the same angular momentum quantum numbers.

The mean square hyperradius of the first excited state is shown in fig. 3 as function of an energy, which is defined as the one- or two-body separation energy above and below the two-body threshold, respectively. Close to the three-body threshold the mean square radius should obey the three-body asymptotics, i.e. go as a logarithm of the three-body binding energy [6]. This occurs outside the range shown in the fig. 3. When the binding energy increases towards about 2 keV, the mean square hyperradius decreases to about $10^4$ fm$^2$. Then both turn around until the excited three-body state disappears into the two-body continuum. The mean square radius has here reached the asymptotic region and scales inversely with energy [6].

**Angular momentum couplings.** We considered a spin zero core and included components of orbital angular momentum less than or equal to 2 in each binary subsystem. In the neutron-neutron subsystem we included $(L_x, L_y, L, S_x, S_y, S) = $ (0,0,0,0,0,0) and (2,2,0,0,0,0), which are compatible with the Pauli principle, $L_x, L_y \leq 2$, positive parity and the total $J = 0$. For the neutron-core subsystem, where $S_x = S_y = 1/2$, we included $(L_x, L_y, L, S_x, S_y, S) = $ (0,0,0,1/2,1/2,0), (1,1,0,1/2,1/2,0), (2,2,0,1/2,1/2,0). The antisymmetric neutron-neutron states were obtained by proper linear combinations. The states with $L = S = 1$ are not included, since they, in the absence of spin-orbit forces, are decoupled from the $L = S = 0$ states. The set of potentials used here, spin independent attractive gaussians, provides binding energies with an accuracy of about 5% by inclusion of only the $L_x = L_y = 0$ components.

When the core has a non-zero spin $S_c$, a number of different structures are possible due to the two angular momenta arising from coupling of the neutron and core spins, which in turn couples to the other neutron spin. An interesting illustrative example is $S_c = 3/2$ with the corresponding neutron-core relative s-state angular momenta $J_{cn} = 1, 2$ and $J = 1/2, 3/2$. The minimum number of components for $J = 3/2$ is then $(L_x, L_y, L, S_x, S_y, S) = (0,0,0,0,3/2,3/2)$ for the neutron-neutron subsystem and $(0,0,0,1,1/2,3/2),(0,0,0,2,1/2,3/2)$ for the neutron-core system. For $J = 1/2$ we need instead at least $(1,1,0,1,3/2,1/2)$ for the neutron-neutron and $(0,0,0,1,1/2,1/2)$ for the neutron-core subsystems. The case of $J = 1/2$, where only $J_{cn} = 1$ enters, necessarily excludes the most favorable configuration of $L_x = L_y = 0$ in the neutron-neutron subsystem.

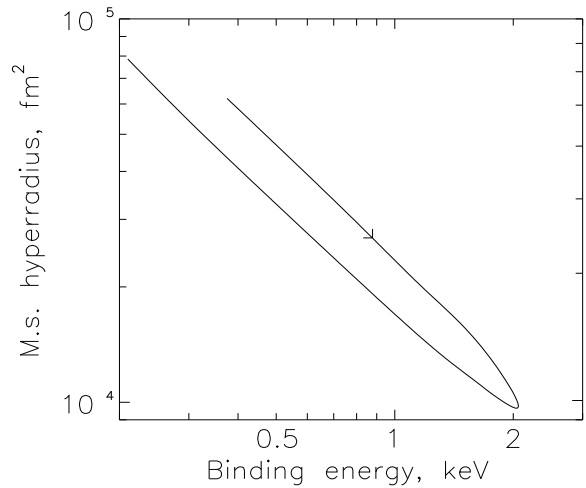

Figure 3: The mean square hyperradius $< \rho^2 >$ of the three-body system as function of an energy defined as the one- or two-neutron separation energy above and below the two-body threshold, respectively. The arrow indicates the direction of increasing potential strength. The core mass is 9 times the nucleon mass.

When $J = 3/2$ both $J_{cn} = 1, 2$ are possible and the neutron-core interaction might depend on the coupling. The two-body threshold could then be split into two separated thresholds. Each would, without the coupling to the other, correspond to the picture exhibited in fig. 2. The inevitable coupling of the $J_{cn} = 1, 2$ states effectively amounts to averaging these two-body states in the three-body solution. Thus Efimov states are hindered, when the difference between the two thresholds is so large that the individual regions of appearance do not overlap, see the dashed line in fig. 2.

**Experimental observation.** It is the neutron-core scattering length which determines the effective radial potential $\lambda(\rho)$ for distances larger that the characteristic radius of the neutron-core potential ($\approx$3 fm). A spatially extended system like the neutron halo nucleus resides mostly in the outer region of the effective potential and is then mainly sensitive to the low energy properties of the two-body potential.

The effect of the unknown particular shape of the neutron-core potential at short distances is larger for the states which have larger probability at short distances. We can thus expect that the binding energy of the ground state might decrease when the potentials with repulsive cores are used. The first exited state having little density at small distances must be less affected. The binding energy of the ground state of the two-neutron halo nucleus, where the Efimov effect might be present, is then expected to be somewhat less than the value of 1 MeV given by the pure attractive gaussian neutron-core potentials.

The investigations have concentrated on a system of two neutrons outside a core nucleus of mass 9. However, for given scattering length in the neutron-core system we find that increasing the core mass essentially leaves the properties of the three-body system unchanged. Thus no special place along the neutron drip-line is favored.

What are then the chances of actually observing an Efimov state in a nucleus? We need to have a neutron s-state, either corresponding to the ground state or an excited state, close to the threshold in the neutron-core system. The first obvious place to look for Efimov states is thus among nuclei with the outer neutrons in the sd-shell. One candidate is $^{14}$Be since recent calculation [9] indicates that the s-level in $^{13}$Be is placed very close to the neutron threshold. The next s-orbitals cross zero energy at mass numbers about 50 and 160 for beta-stable nuclei, but the level density is there so large that it is doubtful whether the core and halo degrees of freedom can be separated. It is only in light nuclei that both the neutron separation energy is small and the level density is low. Even here there is a lack of spectroscopic data, but among the promising systems are e.g. the nuclei $^{17}$C and $^{19}$C that have neutron separation energies of 729±18 and 160±110 keV, respectively, and are expected to have a low-lying $1/2^+$ state. Efimov states might then appear just below the one-neutron thresholds in $^{18}$C and $^{20}$C. The neutron rich oxygen isotopes would also be a possibility, but our knowledge about these nuclei is even smaller at the moment.

**Conclusion.** The neutron halo nuclei are suggested as a new type of system where the Efimov effect might occur. We argue that the most promising place in nuclei to look for the Efimov effect would be in a system, where one neutron is added to a pronounced one-neutron halo nucleus. Our calculations show that the radial extension of the first excited state is very large, at minimum about 100 fm with a corresponding binding energy about 2 keV. This happens when there is a weakly bound state in the neutron-core subsystem corresponding to a scattering length of −20 fm. These properties presumably provide the best conditions for the extremely difficult production of nuclear Efimov states.

**Acknowledgments** One of us (DVF) acknowledges the support from the Danish Research Council.